\documentclass[10pt,twocolumn,letterpaper]{article}

\usepackage[letterpaper,top=0.75in,bottom=1in,left=0.75in,right=0.75in,columnsep=0.33in]{geometry}

\usepackage{times}
\usepackage{helvet}
\usepackage{courier}
\usepackage{amsmath}
\usepackage{amsfonts}
\usepackage{booktabs}
\usepackage{graphicx}
\usepackage{xcolor}
\usepackage[hyphens]{url}
\usepackage[font=small,labelfont=bf,labelsep=colon]{caption}

\usepackage{natbib}
\setcitestyle{authoryear,round,citesep={;},aysep={},yysep={,}}

\usepackage[colorlinks=true,linkcolor=blue!50!black,citecolor=blue!50!black,urlcolor=blue!50!black]{hyperref}
\usepackage[capitalise]{cleveref}

\makeatletter
\gdef\@affiliations{}
\newcommand{\affiliations}[1]{\gdef\@affiliations{#1}}
\renewcommand{\@maketitle}{%
  \newpage\null\vskip 1em%
  \begin{center}%
    {\LARGE\bfseries \@title \par}%
    \vskip 1.2em%
    {\large \@author \par}%
    \vskip 0.8em%
    {\small \@affiliations \par}%
  \end{center}%
  \par\vskip 1.5em}
\renewcommand\section{\@startsection{section}{1}{\z@}%
  {-2.5ex \@plus -1ex \@minus -.2ex}{1.2ex \@plus .2ex}%
  {\normalfont\large\bfseries\centering}}
\renewcommand\subsection{\@startsection{subsection}{2}{\z@}%
  {-2ex \@plus -.5ex \@minus -.2ex}{0.8ex \@plus .2ex}%
  {\normalfont\normalsize\bfseries}}
\makeatother

\newenvironment{links}%
  {\par\smallskip\begin{description}\setlength{\itemsep}{0pt}\small}%
  {\end{description}\par}
\newcommand{\link}[2]{\item[#1:] \url{#2}}

\newcommand{\NUMSPS}{$365{,}842$ }
\newcommand{\NUMNODES}{$2{,}003{,}536$ }
\newcommand{\NUMFILTEREDSPS}{$106{,}424$ }

\newcommand{\NUMFOLLOWS}{846{,}259{,}154}

\newcommand{\NUMFILTEREDPAIRS}{191{,}648}

\title{Finding Common Ground: Graded Communal Knowledge in Bluesky Starter Packs}
\author{
    Sagar Kumar\textsuperscript{\rm 1,2},
    Lawrence Swaminathan Xavier Prince\textsuperscript{\rm 1},
    Julia Mendelsohn\textsuperscript{\rm 3},
    Brooke Foucault Welles\textsuperscript{\rm 1,4},
    Nicholas W. Landry\textsuperscript{\rm 5,6,7}
}
\affiliations{
    \textsuperscript{\rm 1}Network Science Institute, Northeastern University\\
    \textsuperscript{\rm 2}Computational Health Informatics Program, Boston Children's Hospital\\
    \textsuperscript{\rm 3}College of Information, University of Maryland, College Park\\
    \textsuperscript{\rm 4}Department of Communication, Northeastern University\\
    \textsuperscript{\rm 5}Department of Biology, University of Virginia\\
    \textsuperscript{\rm 6}School of Data Science, University of Virginia\\
    \textsuperscript{\rm 7}Vermont Complex Systems Institute, University of Vermont\\
    kumar.sag@northeastern.edu,
    swaminathanxavierp.l@northeastern.edu,
    juliame@umd.edu,
    b.welles@northeastern.edu,
    nicholas.landry@virginia.edu
}

\begin{document}

\maketitle

\begin{abstract}
Communication is made possible by common ground---the unspoken knowledge that people share and presuppose of one another, whether that be online or offline. 
In his conception of common ground, \citet{clark_using_1996} distinguishes between personal and communal common ground, and asserts that the latter is graded: the more community affiliations two people share, the more common ground they share as well.
Social media research has invoked this mechanism to explain how users connect, but it has gone largely untested because community memberships are rarely visible and, where they are, they are coupled to user interactions in a way that leads to conflating effects.
To circumvent these challenges, this study repurposes Bluesky starter packs (SPs) as user-curated community affiliation labels.
Across $\NUMFILTEREDPAIRS$ pairs of users, we show that shared lexical repertoire---our proxy for common ground---grows monotonically with the number of SPs that users share, with users sharing a single pack being roughly twice as similar as equally connected strangers.
A semantic renormalization of SP co-membership shows furthermore that it is more so the number of topically \emph{distinct} communities, rather than the raw count, in which common ground is graded.
Finally, we show that community co-membership adds to common ground independently of proximity in the Bluesky follow network.
These results lead to the conclusion that community membership is a measurable, separable, and semantically structured carrier of common ground. Reading it as such makes common ground observable before an exchange rather than inferred from it, and thus opens the door for large-scale observational approaches to a set of questions that have so far only been posed in the laboratory.
\end{abstract}

\begin{links}
     \link{Code}{https://github.com/sagarkumar16/bsky_common_ground/tree/main}
\end{links}

\section{Introduction}

When two people talk, they rely on a great deal that is left unsaid.
Clark's notion of \emph{communal common ground} holds that much of this unspoken information is inherited rather than negotiated~\citep{clark_definite_1981,clark_using_1996, clark_context_2009}.
Two strangers who establish that they are both cyclists, both Londoners, and both linguists may presume a great deal about one another on the strength of those memberships alone, without ever having met.
Importantly, this inheritance is taken to be \emph{graded}, meaning that sharing more cultural communities should entail sharing more common ground~\citep{clark_using_1996, clark_context_2009}. However, this assumption of gradation by community membership has gone largely untested at scale, as community memberships are rarely visible, and where they are visible, they are not easily measurable.

For this, we look to contemporary social media. The role played by common ground in online communication and social media has often been studied, but rarely named as such. Nevertheless, common ground is often relied upon for the interpretation of texts online, as content posted on a social media platform is generally produced for readers who are presumed to already know a great deal, and writers leave that knowledge unsaid accordingly. Overlooking this dependence on common ground for interpretation has been shown to be an issue for computational approaches to social media text, where private senses can be assigned to ordinary words~\citep{lucy_characterizing_2021} and, therefore, instances of hate speech can be mistaken for innocuous posts~\citep{pavlopoulos_semeval-2021_2021,mendelsohn_dogwhistles_2023}. This is made possible only because, despite there being no prior conversational history, a communal common ground exists in which there is shared knowledge between users from the communities to which they belong~\citep{yus_cyberpragmatics_2011, lampe_familiar_2007}.

Issues can also arise when a reader lacks the requisite common ground to interpret text online. This is made evident and urgent by the findings of~\citet{marwick_i_2011}, in which the authors have shown that when an unintended audience who lacks sufficient common ground gains access to a post on Twitter, it leads to ``context collapse'' and threatens to flatten the social media landscapes by discouraging users from being vulnerable and authentic. This relationship between a text and common ground often appears in the form of community-specific lexicons~\citep{mendelsohn_dogwhistles_2023, zhang_community_2017, lucy_characterizing_2021}. We use this to license the use of \emph{shared repertoire}, in the sense used by~\citet{wenger_communities_1999}, as an observable proxy for communal common ground between users online.

\citet{lampe_familiar_2007} draw directly on~\citet{clark_arenas_1992} to suggest that users online disclose community memberships in their Facebook profiles as a way to establish communal common ground. They note, however, that profile fields are noisy, unverifiable, and unstructured---a limitation that was also acknowledged by~\citet{pathak_method_2021}. A stronger signal would be one conferred by others rather than by the user, specific to a named community, and available in quantity. Social media platforms provide such signals in the form of community memberships (e.g., subreddits, Facebook groups, etc.). However, measuring common ground in these settings is challenging because the rules and norms of these affordances intrinsically shape user experience in a confounding manner~\citep{danescu-niculescu-mizil_mark_2011, baxter_communication_2008, webb_unobtrusive_1966}. This arrangement prevents an understanding of common ground built up \emph{across} shared communities, and makes it difficult to differentiate between the inherited common ground in Clark's conception, and the linguistic accommodation that comes from sharing a communication space~\citep{baxter_communication_2008}. 

To resolve this, we look to Bluesky---an online social media platform that is unique in this regard.
Bluesky supports ``Starter Packs'' (hereafter ``SPs'' or ``starter packs''), which are user-curated lists of accounts, typically organized around a profession, place, identity, or interest \citep{balduf_bootstrapping_2025}.
Users are added to an SP by that SP's creator and cannot add or remove themselves from an SP; they may not even know that they have been added.
The collection of all SPs can be modeled as a hypergraph (also referred to as a \textit{higher-order network}), where each user is a node and every SP is a hyperedge \citep{smith_blue_2026}.
Because the starter pack network is not dyadic, any two users may co-occur in the same SPs any number of times, and the number of SPs that two users share in common may offer insights into how many cultural communities these two users share in common.

In this work, we use starter packs to test Clark's gradation claim, and explore implications for online communication more generally.
Treating starter pack co-membership as an endogenous label for shared cultural community membership, and lexical overlap across users' posting histories as an observable proxy for common ground, we ask whether the latter grows with the former. However, SPs can be noisy and redundant community labels because multiple users may create their own versions of existing SPs. 
To address this, we introduce the ``cross-cutting community count''---an effective count of \emph{distinct} communities that two users are both members of, calculated by semantically renormalizing the number of SPs they share.
Finally, we consider the relationship between common ground and following behavior, as discussed in~\citet{lampe_familiar_2007}. Recognizing that the effect we observe could instead be the consequence of homophily and linguistic accommodation due to social network position~\citep{shalizi_homophily_2011, danescu-niculescu-mizil_mark_2011}, we conclude our analysis by investigating whether our results can be explained by standard notions of social proximity.

\paragraph{Contributions.} Altogether, this study provides:
\begin{enumerate}
    \item Quantitative evidence of communal common ground on a large online platform, and evidence that common ground, as measured by shared repertoire, grows monotonically with the number of communities two users share.
    This is the first direct test, to our knowledge, of Clark's gradation assertion at platform scale.

   \item A mathematical approach to calculating the number of distinct communities two users share on Bluesky, operationalizing \citeauthor{clark_using_1996}'s \citeyear{clark_using_1996} notion of ``cross-cutting'' communities and providing evidence that gradation is relative to this quantity rather than the raw count of shared SPs.
   
   \item Evidence that community overlap and network distance contribute independently to shared repertoire, and that the relative importance of affiliation rises as network proximity vanishes. This suggests a limitation for research on user similarity based on social network topology alone. 
\end{enumerate}

\section{Background and Related Work}

\subsection{Communal Common Ground}

Common ground is a foundational construct in pragmatics, a subfield of linguistics focusing on how context shapes meaning and interpretation. It occupies a central role in coordination-based accounts of communication such as those of~\citet{lewis_convention_1969} and, subsequently, \citet{stalnaker_common_2002,stalnaker_context_2014}.
Within this tradition, common ground refers to the knowledge, beliefs, and suppositions that people share, and which thereby enable successful communication~\citep{clark_using_1996}.

Clark distinguishes two forms of common ground~\citep{clark_definite_1981,clark_using_1996, clark_context_2009}.
\emph{Personal} common ground is accumulated through direct interaction and shared experience, including the incremental process of grounding within a conversation.
In contrast, \emph{communal} common ground derives from shared membership in ``cultural communities'', encompassing the knowledge, assumptions, customs, and lexical resources which individuals possess by virtue of participating in the same social group~\citep{clark_using_1996}. It is communal common ground that our study is concerned with, and specifically Clark's claim that it is \emph{graded}---i.e., it varies in degree rather than being simply present or absent.
Simply put, people participate simultaneously in many communities and infer shared knowledge from these memberships. If two individuals share memberships across several communities, they are expected to possess correspondingly more communal common ground than those who share memberships across only a few.

Shared community memberships may also be ``nested'' or ``cross-cutting''~\citep{clark_using_1996, clark_context_2009}, and we are interested particularly in the latter.
People sharing nested communities share affiliation in increasingly niche groups, whereas people sharing cross-cutting communities share affiliation in a number of topically \textit{distinct} groups.
The assertion that common ground grows with the number of cross-cutting communities would predict that, for example, a cyclist living in London shares more common ground with another cyclist living in London than with a cyclist living in Brussels, because the former pair share two community affiliations rather than one.
This intuition, however, relies on the distinctness of these affiliations.
Capturing this notion of distinctness is central to our measurement strategy.

\subsection{Linguistic Traces of Shared Knowledge}

It is well established that shared knowledge leaves observable traces in language.
Previous work on grounding and reference has shown that people develop increasingly efficient communicative practices as common ground accumulates through interaction, both online~\citep{ yus_cyberpragmatics_2011} and offline~\citep{clark_referring_1986}.
\citeauthor{brennan_conceptual_1996}'s \citeyear{brennan_conceptual_1996} notion of \emph{conceptual pacts} demonstrates that word choices are shaped by shared communicative history and become part of the common ground established between interlocutors.
This is, to our knowledge, the closest existing demonstration that a shared lexical repertoire functions as common ground, and we take it as licensing the general move from lexicon to common ground.
These patterns have also been observed at scale, as work on linguistic accommodation has shown language use to be closely associated with social relationships and interaction histories~\citep{danescu-niculescu-mizil_mark_2011, zhang_community_2017, lucy_characterizing_2021}.

We should be careful, however, about what we take from this literature.
Conceptual pacts are agreements about how to refer to things that emerge between two people over the course of an interaction.
Because they depend on a shared history, they are a feature of the common ground that develops between specific individuals.
Many of the pairs we study, however, have never interacted, and in many cases are not even connected through the follow network.
For that reason, Brennan and Clark's findings do not apply to our setting directly.

Instead, we use their work as the basis for an analogy.
We assume that \citeauthor{wenger_communities_1999}'s \citeyear{wenger_communities_1999} idea of a ``shared repertoire'' stands to community-level common ground in much the same way that conceptual pacts stand to personal common ground.
The key idea is that a conceptual pact reflects a history of interaction: it is a linguistic trace of the knowledge two people have built together.
By the same logic, we argue that a shared repertoire is a linguistic trace of belonging to the same community.

\subsection{Affiliation Networks and Online Communities}

\begin{figure}
    \centering
    \includegraphics[width=0.85\linewidth]{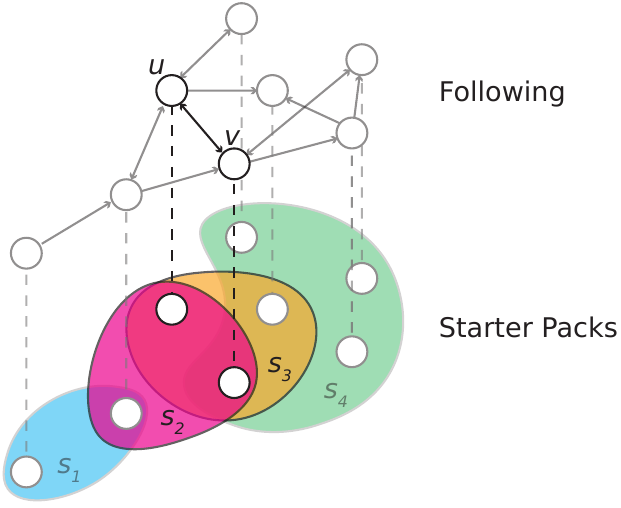}
    \caption{\textbf{The following network and starter pack hypergraph.} Bluesky affords two distinct types of relationships: users following one another, and users sharing the same starter pack. We represent the first as a directed network and the second as a hypergraph in which each SP is a hyperedge comprising its members. In this illustration, dashed lines identify each user across the two representations. Users $u$ and $v$ both follow each other in the following network, so a bidirectional edge is drawn between them in the following network. They are also members of starter packs $s_2$ and $s_3$ but not $s_1$ or $s_4$, each represented as a differently shaded region in the bottom layer. Because $u$ and $v$ share membership in exactly two starter packs, $s_{uv} = 2$.}
    \label{fig:network_layers}
\end{figure}

Research on social networks has long emphasized the role of group memberships in structuring social relationships.
The theory of social foci developed by \citet{feld_focused_1981} argues that ties emerge through participation in shared contexts, while affiliation-network research conceptualizes social structure as arising from the relationships between individuals and the groups to which they belong~\citep{breiger_duality_1974}.
Research on homophily further demonstrates that individuals sharing characteristics or interests are more likely to occupy similar social environments and to form ties~\citep{mcpherson_birds_2001}.

More recently, the field of network science has seen renewed interest in modeling group interactions, often as hyperedges rather than as projections onto pairwise ties~\citep{battiston_networks_2020}. Doing so allows one to change the inferred structure of the system-at-hand~\citep{chodrow_configuration_2020} and offers additional measures of structure~\citep{landry_filtering_2024,kirkley_structural_2025,landry_simpliciality_2024}.
Previous work on Bluesky represented the collection of starter packs as a higher-order network, finding novel structure inaccessible to pairwise representations \citep{smith_blue_2026}.

Empirical work on group membership and interactions online has shown that participation in community structures contributes to the development of shared norms, identities, and linguistic practices~\citep{lampe_slashdot_2004}. Within social media research specifically, common ground has been treated as motivation for users filling out online profiles. \citet{lampe_familiar_2007} draw on Clark to argue that signaling shared community membership through a user profile supplies frames of reference in advance of any interaction. \citet{ellison2011little} read this as suggesting that users seek cues to establish common ground, with the profile lowering the cost of finding them. 
It has not been established, however, whether these factors vary as a function of how many communities two people share, or whether such a relationship survives once network proximity is accounted for. 
These questions are broadly consequential because, more often than not, users online address audiences they cannot see and can only imagine~\citep{marwick_i_2011,litt_imagined_2016}, meaning that who can understand whom on a platform is a question about its communities as much as about its ties. 
As such, these questions bear directly on issues from trust and safety online (e.g., hate speech~\citep{mendelsohn_dogwhistles_2023, elsherief_latent_2021},  marginalization~\citep{marwick_i_2011}, polarization~\citep{chou_public_2025}) to the development of effective strategies for marketing and communication (e.g., target markets and cultural expertise~\citep{grier_noticing_1999, aaker_nontarget_2000}, audience reception~\citep{hall_encoding_2007}).

\subsection{Bluesky Starter Packs}
Bluesky is a microblogging platform similar to X (formerly Twitter), distinguished chiefly by the degree of control it affords users over content discovery and feed curation.
One mechanism supporting this is the ``Starter Pack'': a user-generated collection of between 8 and 150 accounts which may be followed simultaneously.
Starter packs are frequently organized around shared interests, professions, locations, identities, or activities.
In this study, we interpret membership in an SP as a form of endogenous community attribution, where users have been collectively identified as belonging to a particular social context.
Because membership is conferred by another user rather than self-declared, we posit that it is closer to the more robust class of signals called ``assessment signals'', compared to other potential indicators such as users' profiles, which constitute the weaker form of ``conventional signals''~\citep{donath_signals_2007, lampe_familiar_2007}.

We study starter packs as opposed to subreddits or other similar mechanisms because starter packs offer community labels, but are not sites of communication.
There is no such thing as ``posting to a starter pack''---it is simply an attribution that one user gives to another.
Subreddits and forums, on the other hand, are online communities in themselves, acting as sites of communication and governance in ways that couple membership to norms and utterances.
For example, a post in \texttt{r/cycling} is a post \emph{addressed to} r/cycling, and its language reflects that audience.
Speakers design their utterances for their addressees~\citep{bell_language_1984, clark_hearers_1982, baxter_communication_2008}, and communities of practice develop linguistic styles to which participants accommodate upon entry~\citep{eckert_language_2000,danescu-niculescu-mizil_mark_2011}.
Lexical overlap between two r/cycling posters, measured on their r/cycling posts, therefore measures the subreddit's specific style at least as much as it measures anything which the two posters carry with them.
Because starter packs do not provide a forum for posting, we do not need to differentiate between content posted to a starter pack and posts to a user's timeline, and can simply ask what a user's community membership predicts about their complete posting behavior.

\section{Methods}

\subsection{Network Analysis}

We use both pairwise and higher-order network data.
The following network can be represented as a pairwise, directed network, $G = (V,L)$, where $V$ is the set of users, represented as nodes, and $L$ is the set of dyadic edges such that each edge $(i, j)\in L$ indicates that ``user $i$ follows user $j$''.
Based on platform restrictions, the resulting network contains neither self-loops nor multi-edges.

We represent the collection of SPs as a higher-order network, $H=(V, E)$, where, as before, $V$ is the set of $N=|V|$ nodes but now, $E$ is the set of $M=|E|$ hyperedges, where each SP is a hyperedge of order $m$ (i.e., consisting of $m$ users).
Because starter packs do not rely on follows and co-memberships are not dyadic, a pair of users can co-occur in any number of starter packs despite no multiedges existing in the following network.

We can also represent the higher-order starter pack network as an incidence matrix $B \in \{0, 1\}^{N \times M}$, where $B_{ij} = 1$ if node $i$ belongs to hyperedge $j$.
Let $\mathcal{P} = V \times V$ be the set of all pairs and let $E_i$ be the set of all hyperedges (i.e., SPs) that node $i$ is a member of.
For a pair of distinct users, $u, v \in \mathcal{P}, u \neq v$, we can then define the count of overlapping SPs as
\begin{equation}
s_{uv} \;=\; |E_u \cap E_v| \;=\; \sum_{e \in E} B_{ue} B_{ve}
\;=\; \left(B B^T\right)_{uv}.
\label{eq:shared}
\end{equation}
Similarly, a node's degree (i.e., the number of SPs to which she belongs), is defined as $k_u = \sum_e B_{ue}$.
Finally, we define network distance $\ell_{uv}$ as the length of the shortest path between $u$ and $v$ in the directed Bluesky follow graph, $G$.

\subsection{Data}
\label{sec:data}

All data that were used come from~\citet{smith_blue_2026}, released under the CC BY 4.0 License. This paper details the process by which collected from Bluesky’s Sync API. 
The data are structured as a JSON Lines file for each active user, storing the user ``repository'', which contains all public user actions.
The data used in this study were the starter pack network, the following network, and all textual posts (posts, replies, and quote posts).
All users referenced in the data by a universal decentralized identifier (DID).

\subsubsection{Starter pack network}

As detailed in~\citet{smith_blue_2026}, all starter packs from all user repositories---comprising \NUMSPS total starter packs containing a total of \NUMNODES user accounts---were collated.
In addition to memberships, we also stored each starter pack's title and description.

\subsubsection{Following network}

We collated the entire following network, and used the network induced on the accounts specified by the filtered starter pack hypergraph in subsequent analysis.
Although this network is temporal (in that we have the timestamp at which a follow occurs), we only considered the static following network for simplicity.
This network contains \NUMNODES accounts with \NUMFOLLOWS~directed following relationships.
and is quite heterogeneous in both its in- and out-degree \citep{smith_blue_2026}.

\subsubsection{User posts}
For each user in the filtered starter pack network, we collected their 1,000 most recent posts, replies, and quote posts, to form a corpus of text for each user.
The Bluesky platform's official account was identified and removed from the set of accounts each user follows, as it is followed automatically by every new user and thereby distorts all degree-dependent quantities.
Bluesky is a multilingual platform~\citep{balduf_bootstrapping_2025}, but because our similarity measure operates on a shared vocabulary, cross-language comparisons would register as dissimilarity regardless of common ground.
We therefore restrict the corpus to posts written in English, and return to the implications of this restriction in the Limitations and Ethical Considerations sections.

\paragraph{Text preprocessing}
Because we measure common ground through content rather than style, preprocessing was designed to isolate the content signal.
Each post was lowercased and tokenized, and URLs, user mentions, standard platform boilerplate, and stopwords were all removed. The remaining words were reduced to their base forms.
Hashtags were retained as tokens as they frequently act as community markers independent of the words they contain~\citep{zappavigna_ambient_2011}.
Pairs in which either user falls below $50$ usable tokens after preprocessing were dropped because stable lexical representations cannot be constructed for them. Following this filtration, a total of $\NUMFILTEREDPAIRS$ pairs remained, spanning $70{,}575$ unique users, and \NUMFILTEREDSPS SPs.

\subsection{Sampling and Uncertainty}
\label{sec:sampling}

\begin{figure}[t]
    \centering
    \includegraphics[width=\linewidth]{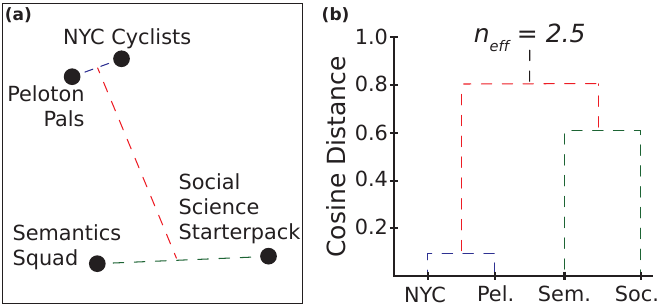}
    \caption{\textbf{An illustration of the effective shared community calculation.}
    \textbf{(a)} An illustration of four hypothetical starter packs shared by users $u$ and $v$ embedded in $n$-dimensional latent space.
    Here, the cycling-related SPs are semantically similar, and the other two are less similar.
   \textbf{(b)} An illustration of the measure described in Eq.~\eqref{eq:neff-int}.
    The number of surviving clusters $c(\tau)$ is integrated over all thresholds, $\tau$, yielding the effective count $s^{R}_{uv}$.}
    \label{fig:renorm_sketch}
\end{figure}

\paragraph{Sampling} 
Because analyzing all $\mathcal{O}(10^{10})$ possible pairs in the dataset is computationally infeasible and pairs with a large number of shared SPs are rare, we sampled by conditioning on the number of SPs two users share.
Uniformly sampling pairs of users would over-represent the low-overlap regime and leave the high-overlap regime under-represented, so we exploit the fact that the number of starter packs shared by users $u$ and $v$ can be directly calculated using the incidence matrix, as described in Eq.~\eqref{eq:shared}. With this, we grouped all pairs by $s$, denoting this collection of pairs $\mathcal{P}_s$.
We then drew $75{,}000$ samples from $\mathcal{P}_s$ for each $s \in \{1, 2, \dots,8\}$.
For the $s=0$ case, however, there are far too many pairs to sample directly and we instead used rejection sampling, drawing a pair uniformly at random and rejecting it if $s>0$ until we had reached $150{,}000$ samples.
To address the distributional differences between these two sampling methods, we introduced a degree-matched baseline which is motivated and detailed in Appendix A.

\paragraph{Uncertainty under dyadic dependence.}
Because each user participates in many pairs, similarities involving the same user are correlated, and naive standard errors, or bootstraps that resample pairs, understate uncertainty~\citep{aronow_cluster_2017}.
We therefore employ a user-level cluster bootstrap.
For bootstrap sample $c$, users were sampled with replacement. Let $m_u^{(c)}$ be the number of times user $u$ is selected, then the corresponding weight for pair $(u,v)$ is $w_{uv}^{(c)} = m_u^{(c)} m_v^{(c)}$~\citep{aronow_cluster_2017,davezies_empirical_2021}. We report 95\% confidence intervals based on 500 bootstrap samples throughout.

\subsection{Estimating Common Ground}
\label{sec:sim}

As discussed above, we leverage shared lexical repertoire as a proxy for common ground. Operationalizing this requires a measure which fits the following criteria.
If communal common ground manifests (at least in part) in community-supplied words, then (1) the measure must weight community-specific terms
heavily and platform-general terms very little.
It must also (2) be symmetric, as common ground is shared between interlocutors.
Finally, it should (3) be well-defined for pairs sharing almost no vocabulary and (4) inexpensive to evaluate.

These requirements lead us to estimate common ground by calculating the cosine similarity between $L_2$-normalized TF-IDF vectors~\citep{manning_introduction_2008}, ${\bf x}_u = \left( x_{t_1, u}, x_{t_2, u}, \dots \right)^T$, for each user $u$ and term $t$. 
We employed sublinear term frequency~\citep{salton_termweighting_1988} to ensure that a term repeated many times by a prolific user does not dominate that user's vector, together with smoothed inverse document frequency in \texttt{scikit-learn}~\citep{pedregosa_scikitlearn_2011}.
We also pruned the vocabulary at both ends so that terms appearing less than $10\%$ of user timelines were removed to prevent typos from inflating similarities, and terms appearing in more than $40\%$ of user timelines were removed to lessen the effect of general platform-wide lexical overlap. For results corresponding to other thresholds, see Appendix B. Altogether, this left us with a lexicon, $\mathcal{T}$. 
For a detailed discussion regarding why we selected TF-IDF instead of pretrained embeddings from a large language model, see Appendix C.

\begin{figure}
    \centering
    \includegraphics[width=\linewidth]{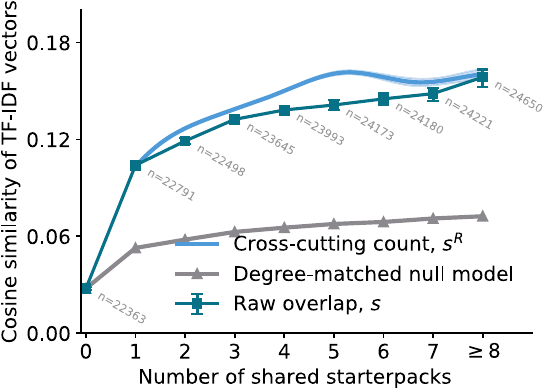}
    \caption{\textbf{Common ground grows and saturates quickly with respect to the number of shared starter packs.}
    Mean cosine similarity with respect to the raw number of shared starter packs ($s$; square markers) and the cross-cutting community count ($s^R$; solid line fit using a GAM) of SPs that two users share, plotted against the $s=0$ null (triangle markers) with $95\%$ cluster-bootstrap confidence intervals. Each value $s \in \{0, \dots, 7, \geq 8\}$ annotated with the sample size. The 
    }
    \label{fig:count_v_sim}
\end{figure}

\subsection{Semantic Renormalization of Shared Starter Packs}
\label{sec:semantic}

\begin{figure*}
    \centering
    \includegraphics[width=.8\linewidth]{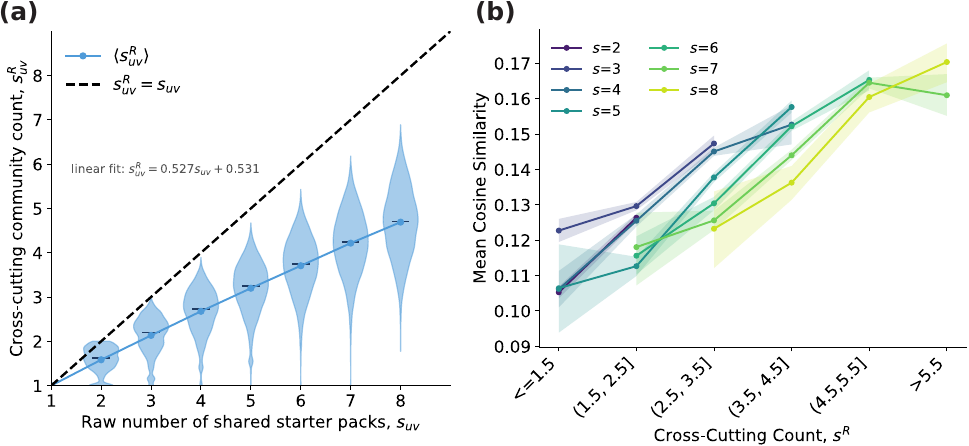}
    \caption{\textbf{Shared SPs are substantially redundant, and distinctness predicts similarity.} \textbf{(a)} The distribution of cross-cutting community count as a violin plot with respect to the raw number of shared SPs. The dashed black line illustrates when the number of cross-cutting communities matches the raw number of shared SPs.
    The solid blue line shows the estimated linear best-fit line fit to the mean value of $s^R$ with respect to $s$, which demonstrates that the number of cross-cutting communities is, on average, roughly half the number of raw SPs. \textbf{(b)} Mean cosine similarity as a function of the cross-cutting count $s^{R}$, with pairs stratified by $s$, their raw count of shared SPs, alongside $95\%$ cluster-bootstrap intervals. Curves are truncated at $s^{R} \le s$ (by construction) and pooled above $s^{R} = 5.5$ where cells become sparse.}
    \label{fig:crosscutting}
\end{figure*}

Raw co-membership count does not account for similarity between communities.
Consider a pair sharing five SPs, four of which are related to the same topic. Consider also a second pair sharing three SPs on entirely distinct topics.
By Clark's cross-cutting logic, the latter pair ought to possess more communal common ground, despite the smaller raw count.
We therefore introduce a ``cross-cutting community count'', $s^R$, which counts \emph{distinct} communities by discounting semantic redundancy among the SPs a pair shares.

This is done by renormalizing $s_{uv}$ for any given pair, $(u,v)$, based on the names and descriptions of the starter packs they share. 
For each SP, we concatenated the name and description and mapped this to a $384$-dimensional unit-norm sentence embedding using MiniLM-L6-v2 from \texttt{sentence-transformers}~\citep{reimers_sentencebert_2019}. This yielded a pairwise cosine distance matrix over all SPs.
For each pair, we then grouped all shared SPs according to their similarity using average-linkage hierarchical clustering~\citep{manning_introduction_2008}.

If we let $c(\tau)$ denote the number of clusters at similarity threshold $\tau \in [0,1]$, then at $\tau=0$, every SP is its own community and $c(0) = s_{uv}$.
As $\tau$ grows, communities merge, and when $\tau = 1$, a single community remains.
Because we have no justification for selecting one threshold over another, we employ a standard technique in topological data analysis and integrate over all scales~\citep{carlsson_topology_2009}, yielding a renormalized count
\begin{equation}
s^{R}_{uv} \;=\; \int_{0}^{1} c(\tau)\, \mathrm{d}\tau=1 + \sum_{k=1}^{s_{uv}-1} h_k.
\label{eq:neff-int}
\end{equation}
where $h_1,\dots,h_{s_{uv}-1}$ are the $\tau$ values of the merges (as illustrated in Fig.~\ref{fig:renorm_sketch}(b)).
This bounds our ``cross-cutting community count'', $s^{R}_{uv}$, such that $s^{R}_{uv} \in [1, s_{uv}]$. Here, the lower bound is attained when all shared SPs are semantically identical and the upper bound is the case when all SPs in $E_u \cap E_v$ are mutually orthogonal.

It should be noted that while this construction discounts redundant SPs, it also consequently discounts nested communities, e.g., ``NYC Cyclists'' and ``Brooklyn Cyclists''.
However, we are only considering the assertion of graded common ground for cross-cutting communities, and not Clark's parallel assertion for nested communities, making this artifact a matter for future studies.
Nevertheless, as raw counts inversely \emph{over}count communities, we report both throughout, and posit that the exact relationship between overlapping cultural communities and common ground sits somewhere in between these two cases. 

\subsection{Network Distance}
\label{sec:network}

Positive results coming from these analyses can provide evidence towards Clark's assertion, but they cannot rule out the possibility that SP co-membership may simply be a proxy for social proximity because starter packs drive follows, follows drive exposure, and exposure drives similarity~\citep{shalizi_homophily_2011, danescu-niculescu-mizil_mark_2011, lampe_familiar_2007}.
If this were true, $\mathrm{sim}(u,v)$ would be well-predicted by the network proximity, and community overlap would be a random effect.

To address this, we obtained the following distance between users in each pair using breadth-first search on $G$, limited to paths of length $\ell \leq 3$.
We then grouped all pairs by the number of effective shared starter packs into three bins, $s^R=0$, $1\leq s^R\leq 3$, and $s^R >3$ and compared how similarity varied across network distance, as well as across starter pack co-membership. 

\subsection{Null Model}
\label{sec:null}

Because users in many starter packs are more visible and more active~\citep{smith_blue_2026}, high-activity users may converge upon platform-wide language use which would inflate similarity among pairs which also have high $s_{uv}$, and a positive relationship between $s_{uv}$ and $\mathrm{sim}(u,v)$ might arise despite no direct effect from cultural communities. To address this, observations must be compared against a null model. 

A natural null model for SP membership would be a bipartite (or hypergraph) configuration model.
However, because our analysis requires pair sampling based on $s$, a user $u$ in $k_u$ SPs is more likely than a user $v$ in $k_v$ SPs to be sampled for any given value of $s$ if $k_u > k_v$.
Therefore, using the hypergraph configuration model as a null model can cause misleading artifacts (see Appendix A).

To address this, we instead developed a degree-matched null model.
For each observed pair $(u,v)$ sharing $s$ SPs, we recorded the joint degree profile $(k_u, k_v)$, binned degrees into deciles, and sampled zero-overlap (i.e., $s = 0$) pairs to replicate the distribution in each observed set of $\mathcal{P}_s$ pairs.
The comparison set thus differs from the observed pairs only in whether the users share SPs, and not in how many packs they belong to (and consequently, how much they post).
Matching is performed upon $k_u$, while matched pairs are drawn from the same vectorized subpopulation.
We thus define the degree-controlled effect of a starter pack co-membership for a bin $s$ as
\begin{equation}
    \Delta(s) \;=\; \langle \mathrm{sim}_s\rangle - \langle \widetilde{\mathrm{sim}_s}\rangle, 
\label{eq:delta}
\end{equation}
where we define $\langle \widetilde{\mathrm{sim}_s}\rangle = \frac{1}{|\tilde{\mathcal{P}}_s|}\sum_{(u^\prime,v^\prime) \in \tilde{\mathcal{P}}_s} \mathrm{sim}(u^\prime,v^\prime)$ and $\tilde{\mathcal{P}}_s$ denotes the set of degree-matched zero-overlap pairs constructed for bin $s$.

\section{Results}

\subsection{Common Ground Grows With Co-membership}

Plotting mean cosine similarity, $\langle \mathrm{sim}_s \rangle$, against the raw count of shared starter packs, $s$, we observe that the cosine similarity rises monotonically, from $0.028$ at $s=0$ to $0.159$ at $s \ge 8$, with means separated well beyond their confidence intervals across most of the range (Fig.~\ref{fig:count_v_sim}). This value also considerably exceeds the growth in similarity that we see from the degree-matched baseline through. 
The shape of the growth is itself also informative.
The step from $s=0$ to $s=1$ is by far the largest, after which growth continues at a markedly lower and roughly constant marginal rate, as fitting over $s \in [1,8]$ gives $\beta = 6.7 \times 10^{-3}$ $[6.1, 7.3] \times 10^{-3}$ per additional pack.
Extending the raw count before pooling suggests that this trend appears stable as we increase $s$, as we assume that this continues into the long tail of the SP co-membership distribution (see Appendix F).
Given decaying sample size and statistical power at large $s$, however, we focus the rest of our analyses on this early-stage saturation and pool results at $s \ge 8$.

\subsection{Cross-Cutting Communities Count for More}
\label{sec:seff}

Fig.~\ref{fig:count_v_sim} also shows mean cosine similarity as a function of the renormalized, cross-cutting count $s^{R}$, fit to a generalized additive model. 
This is compared against the raw count of SPs, and the two coincide at a value of one (where $s^{R} = s$ by construction). Values diverge immediately afterwards, with the $s^{R}$ curve lying above the raw curve throughout.
This upward displacement indicates that similarity rises more steeply per \emph{distinct} than per raw SP. This suggests the topical redundancy in starter packs does not contribute as much as ``cross-cutting'' community co-memberships. 

Across $s$, renormalization reduces co-membership numbers substantially. 
Fig.~\ref{fig:crosscutting}a shows the distribution of cross-cutting community count, $s^{R}_{uv}$, for each raw count of SPs. It is clear that the mean cross-cutting count falls well below equivalence for every value of $s$. 
The gap appears to widen with $s_{uv}$, so that pairs sharing eight SPs share, on average, fewer than five distinct, cross-cutting communities. 
Raw co-membership therefore overstates cross-cutting community co-membership for the large majority of pairs, and does so increasingly as the raw count grows.

Examining the effect of distinctness versus redundancy within each level of $s$, Fig.~\ref{fig:crosscutting}b plots
mean cosine similarity as a function of $s^R$, separately for each raw count $s \in \{2, \dots, 8\}$, with $95\%$ cluster-bootstrap intervals.
($s{=}1$ is excluded because $s^R \equiv s = 1$ there; we pool at $s^{R} > 5.5$, beyond
which sample sizes become too sparse.) Reading a single curve from left to right moves from
pairs with more redundancy in their shared SPs to pairs with more cross-cutting community co-memberships (i.e., more topically distinct shared SPs), with the raw count held fixed. Every curve rises, meaning that at every
value of $s$, mean similarity increases with $s^{R}$ monotonically within uncertainty. Moreover, the magnitude of the within-$s$ rise, moving from the
least to the most distinct configurations raises mean similarity by roughly $0.05$, which is about as much as the entire rise from $s{=}1$ to $s{\ge}8$ in Fig.~\ref{fig:count_v_sim}. This gradient remains consistent for both topical and person-centric pack types (see Appendix D). Altogether, these results confirm that, at identical raw counts, pairs whose shared SPs name more distinct, cross-cutting communities are consistently more similar than pairs with semantically redundant SPs.
 
\subsection{Shared Communities Contribute Beyond Proximity}
\label{sec:netresults}

\begin{figure}
    \centering
    \includegraphics[width=\linewidth]{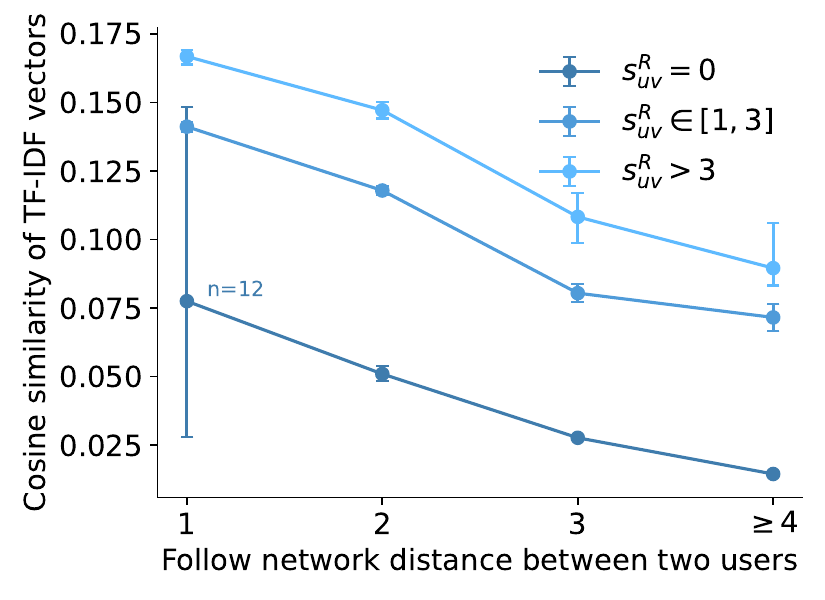}
    \caption{\textbf{Community overlap and network distance contribute independently to common ground.} Here, we plot the mean cosine similarity with respect to the network distance via the following network, stratified by effective SP co-membership, with $95\%$ cluster-bootstrap intervals. The $s^{R}{=}0$ point at $\ell{=}1$ rests upon only $12$ pairs and is not interpreted.}
    \label{fig:network_comparison}
\end{figure}

Finally, we ask whether co-membership in communities contributes anything beyond what can be explained by position in the following network, or if it is merely a proxy for social proximity.
Fig.~\ref{fig:network_comparison} shows that within the highest-overlap stratum (selected by rounding the median of the $s^R$ distribution), mean similarity falls monotonically from $0.167$
$[0.164, 0.169]$ at $\ell = 1$ to $0.090$ $[0.083, 0.106]$ in the $\ell > 3$ bin---a decline of $46\%$, with disjoint intervals at every consecutive step. 

Despite the strong effect of network distance, Fig.~\ref{fig:network_comparison} also shows clear separation and stable ordering between $s^R$ strata.
Moreover, the effect of community co-membership is roughly constant across all network distances, as measured in the difference in similarity between each of these $s^R$ strata.

\section{Discussion}

In this study, we tested the assertion that communal common ground is graded by the number of cultural communities two people share \citep{clark_definite_1981, clark_using_1996, clark_context_2009}.
We operationalized cultural community co-membership using Bluesky starter packs and common ground as shared lexical repertoire, and asked whether the latter grows with the former.
We found that not only does common ground increase monotonically with SP co-membership, but that it exceeds what might be expected at random for any number of shared SPs.
To our knowledge, this constitutes the first direct evidence for the gradation claim at platform scale.

Our comparison also renders the claim more precise than Clark states it, in two respects.
The first concerns the precise relationship between shared communities and common ground.
We find that this effect quickly saturates; two-thirds of the increase in similarity occurs by the third shared community, and pairs sharing even one SP are roughly twice as similar as equally connected pairs without a shared SP.
We therefore suggest that sharing any cultural communities whatsoever qualitatively changes the relationship between users online,  but additional shared communities yield diminishing returns on similarity.
The second respect concerns what is being counted.
Holding the raw count fixed and varying the cross-cutting community count, we find that pairs whose shared SPs name distinct communities are consistently more similar than pairs that share redundant starter packs, at every level of $s$.
This suggests that it is the number of distinct communities, rather than the raw number of shared SP attributions in which common ground appears to be graded.

Going beyond Clark, we address the argument that co-membership in a starter pack or cultural community is merely a proxy for social proximity, and we find that this is not the case.
Though network distance clearly impacts common ground, community overlap contributes a stable additive increment of approximately $0.027$ across distances.
We thus social proximity on the internet and community affiliation as pproximately additive contributions, meaning that while both mechanisms are correlated with common ground, community affiliation acts even in the absence of any meaningful social network paths between users.

We take these results as providing evidence that communal common ground exists online and is graded by cross-cutting communities, and we consider three implications of this finding for the study of social media. 
The first concerns network proximity.
Distance in the follow graph has long been understood as a signal of similarity between users~\citep{mcpherson_birds_2001, feld_focused_1981}, but our results show that this premise leaves something systematic behind:
at every distance we measure, including among pairs with no meaningful paths between them, users sharing more distinct communities remain more similar. 
This suggests that network distance on community-oriented platforms should be understood as just one feature which determines user similarity, which can be modulated by shared community affiliations that may be invisible to the following network. 
The second implication for social media scholarship is methodological. 
Observational research has long wanted community labels that do not interfere with the behavior under study~\citep{webb_unobtrusive_1966}. 
However, the available constructs either require self-declaration, which is sparse and performative, or derive membership from participation in a venue, which couples the label to the very language being measured. 
Starter Packs are attributions made \emph{about} users, frequently without their involvement, attached to no venue in which anything is posted---in effect, community labels without a community register---and we suggest that structures of this kind, wherever platforms provide them, resolve much of this longstanding difficulty. 
Finally, our findings support the claim that communal common ground itself deserves a place among the working constructs of social media research. 
This study provides methods to measure it at scale, and evidence to suggest both that it behaves as the theory suggests (e.g., that it is graded and sensitive to topical distinctness) and that there are new findings to be made (e.g., that it is saturating and distinct from social proximity). Moreover, it is a construct with immediate applicability wherever anticipating the common ground of strangers has value: audience design, cold-start recommendation, the segmentation of publics for outreach or marketing, etc. The finding that the unspoken presumptions between two people who have never met can be estimated from community attributions alone is, we think, the finding with the
greatest reach.

From these considerations, several potential applications follow, which we offer as directions rather than results. For example, context collapse, which has been characterized primarily through interview and diary studies~\citep{marwick_i_2011, litt_imagined_2016}, could be measured as the variance of common ground across a post's realized audience. Two narrower applications follow from the same logic. On one hand, engagement-based ranking optimizes for whether a viewer will interact with a post, based on the viewer's own past interactions and those of their immediate social network, but not for whether the viewer shares the common ground the post presupposes. A graded estimate of communal common ground could therefore offer an additional term to that optimization problem, introducing information about the cultural background a viewer brings to a post---information that, as our results show, is carried by community affiliation and is not recoverable from network position or interaction history alone. On the other, this kind of information regarding how much relevant community-derived knowledge is shared by users posting deceptively harmful content and the moderators, annotators, and classifiers that are tasked with labeling it as such could be highly beneficial for applications in online trust and safety~\citep{sap_annotators_2022, elsherief_latent_2021}.

While these applications offer enticing directions, it is important to acknowledge that our results come with several important limitations, which we describe here.
First, common ground is far richer than lexical content overlap.
We suggest that our framework measures shared repertoire in Wenger's sense, but the necessary caveat is that it does not capture other plausible ways of establishing common ground in text, such as adopting similar linguistic styles, matching levels of formality, or speaking in languages besides English.
Moreover, we employ a bag-of-words representation which makes our analysis blind to word order, longer constructions, and polysemy.
However, we note that its insensitivity to paraphrase is, for our construct, a requirement rather than a defect (see Appendix C).
Second, our renormalization technique discounts nested communities, which Clark treats as genuine sources of common ground, and includes SPs that were created automatically using the, ``Make one for me'' feature on Bluesky. As such, our two counts ($s$ and $s^R$) are more likely to be bounds on (as opposed to precise measurements of) the true value for shared communities between users.
Finally, for computational feasibility, a number of concessions had to be made:
our sample consisted of only a small fraction of disproportionately active users, our degree matching is decile-coarse, and the long tail of $s$ was left broadly unexplored.

The directions forward follow from these limitations.
Further analysis employing stylistic rather than content features would test whether the structure we observe is a property of common ground, more broadly construed than only TF-IDF cosine similarity.
Furthermore, a membership-keyed renormalization could credit nested communities rather than discounting them.
Theoretically, we also suggest that the relation between shared repertoire, field of discourse, and common ground deserves a robust treatment in its own right.
More broadly, we hope this encourages theoretically motivated empirical work on the pragmatics of online social networks, where emergent platforms now support those structures which theories of language use have long had to assume rather than observe.

\section{Acknowledgments}

Nicholas Landry acknowledges support from the University of Virginia Prominence-to-Preeminence (P2PE) STEM Targeted Initiatives Fund, SIF176A Contagion Science.
The authors acknowledge Research Computing (rc.virginia.edu) at The University of Virginia for providing computational resources and technical support that contributed to the results reported in this publication.

\section{Ethical Considerations}
\paragraph{Data and PII Considerations.} We work entirely from a previously collected, deidentified, and published dataset, and research questions were scoped so that they could be answered without collecting any further data. Our work required no profile
attributes, demographic inference, or authenticated content. We likewise limited computation to what the question demanded, capping post histories and sampling pairs rather than evaluating the full census. All analyses were run on CPU nodes of an internal university cluster, with the total cost being approximately one week of core-walltime.

We report only aggregate, population-level quantities, and release no user identifiers, Pack names, or user text, as verbatim posts are readily re-identifiable through search even when identifiers are stripped,
and releasing unfiltered social media text could potentially expose readers to offensive material. All code sufficient to reproduce our analyses is available at the link provided above.

\paragraph{Potential for misuse.} Methods for inferring community boundaries from public affiliation data could be repurposed for audience segmentation or targeting in ways the observed users did not anticipate. 
Furthermore, as Starter Pack membership is a third-party attribution, a user's Packs may reveal affiliations they did not themselves disclose. Relatedly, $s^{R}$ measures how distinctively positioned a pair is across communities, and could be inverted to identify unusually well-bridged accounts. To mitigate these issues, all users remain deidentified, and we report no individual-level values and no rankings of users or Packs.

\paragraph{Scope.} Although Bluesky is multilingual, we restrict to posts identified as English to prevent additional computational hurdles. As such, the communities
our measures can see are correspondingly a subset of those present. Despite this, we have attempted to develop the research question and methods in an expansive and inclusive manner, rather than restrictive and exclusive. This is reflected in our questions which focus on what community membership predicts about language in general, rather than testing a hypothesis about any particular community or category of user, and the fact that we make no claims about groups our sample under-represents.


\appendix

\section{Appendix A: Why the Configuration Model Manufactures a Threshold}
\label{app:sizebias}

Here we motivate our degree-matched null model by making evident the biases that emerge from the hypergraph configuration model in our experimental setup, and showing how our null model removes these biases.
In short, a configuration model applied to an affiliation hypergraph destroys co-membership by construction, making edge overlap between users a function of joint degree alone.
Conditioning on this means that for an observable of interest that is degree-dependent, a plot of that observable against edge overlap in the configuration null remains heavily biased.
In the study at hand, this causes an apparent threshold (see Fig.~\ref{fig:bad_null}) which ought not to be interpreted, as it is an artifact of the null model.

\subsection{Setup}

\begin{figure*}
    \centering
    \includegraphics[width=\textwidth]{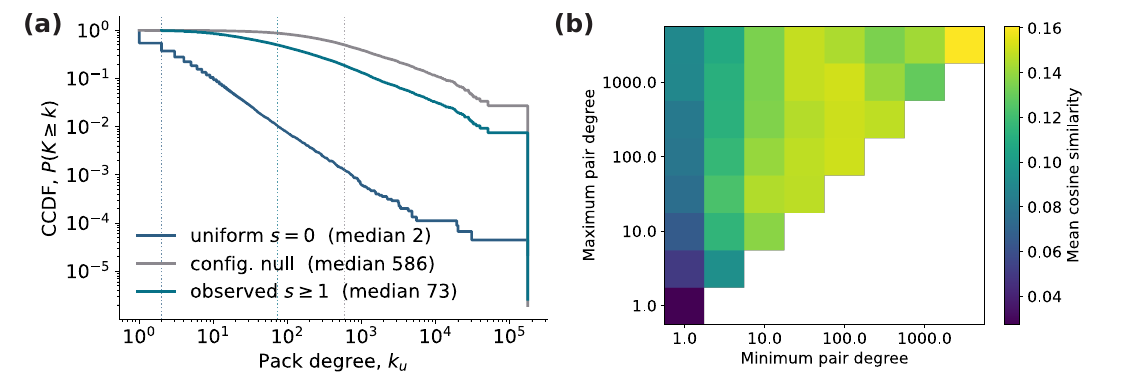}
   \caption{\textbf{The observed and null curves are drawn from different populations, and degree predicts similarity on its own.} \textbf{(a)} Complementary CDF of degree, $k_u$, among the users in the uniformly sampled $s=0$ population, the configuration null produced by the \texttt{bicm} package, and the $s \ge 1$ pairs observed in the data. Dotted lines mark medians.  \textbf{(b)} Mean cosine similarity among \emph{disjoint} pairs ($s_{uv}=0$), binned by the pair's minimum and maximum degree.}
   \label{fig:degree_dists}
\end{figure*}

Prior work~\citep{smith_blue_2026} has established that inclusion in a starter pack drives follower acquisition and posting activity. This was originally a finding about platform dynamics but in this study, we also now recognize it as the premise of a methodological hazard.

Posting activity determines how many tokens survive filtering (see Appendix B), which determines the density of a
user $u$'s lexical vector and thus, in part, determines their cosine similarity to an arbitrary partner, $v$, regardless of $s_{uv}$. We thus introduce
\begin{equation}
g(k, k') \;:=\; \mathbb{E}\big[\,\mathrm{sim}(u,v) \,\big|\, k_u = k,\; k_v = k',\; s_{uv}=0 \,\big]
\label{eq:g}
\end{equation}
as the expected similarity as a function of degree alone between users who have no shared starter packs.
The empirical claim is that $g$ is increasing with $\max(k, k')$, and this is verified in Fig.~\ref{fig:degree_dists}. Two prolific strangers are more lexically similar than two sparse strangers, purely because at least one user has a denser lexical vector. Any estimator that fails to hold the joint degree $(k_u,k_v)$ fixed will therefore confound the co-membership effect with variation in $g$.

The reasoning for this comes from the mechanics of configuration models. Under a canonical hypergraph configuration model, each membership is an independent Bernoulli with 
\begin{equation}
    p_{ue} = x_u y_e/(1 + x_u y_e)
    \label{eq:bernoulli_prob}
\end{equation}
for fitted Lagrange multipliers, $x_u, y_e$, so the reshuffled co-memberships, $\tilde{s}_{uv}$, follow a Poisson-binomial with mean
\begin{equation}
\lambda_{uv} \;=\; \mathbb{E}\big[\tilde{s}_{uv}\big] \;=\; \sum_{e \in \mathcal{E}} p_{ue}\, p_{ve}.
\label{eq:lambda}
\end{equation} 

\begin{figure*}
    \centering
    \includegraphics[width=\linewidth]{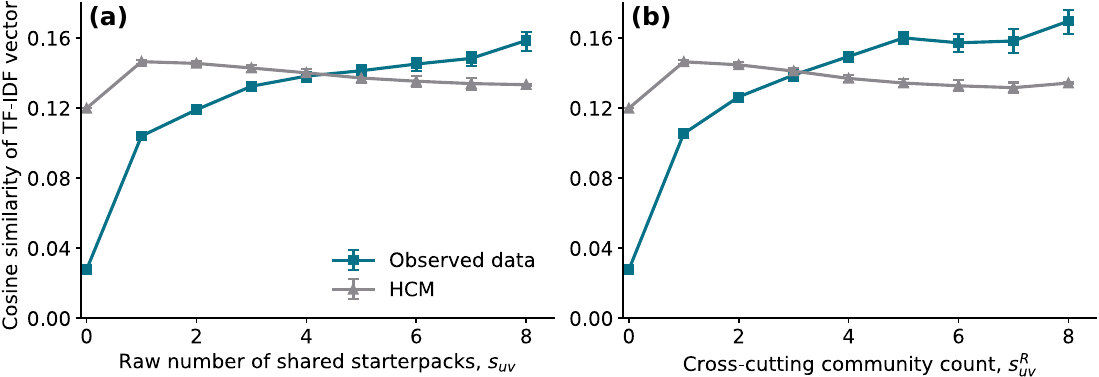}
    \caption{\textbf{The hypergraph configuration model artificially inflates shared repertoire.} Observed similarity between users (square markers) is lower than the configuration null (triangle markers) when $s<3$, implying that sharing a community leads to lower shared repertoire than expected with respect to a sample from the hypergraph configuration model. This is not the case, and is merely an artifact of the configuration model when comparing a degree-correlated observable with pairwise edge-overlap.}
    \label{fig:bad_null}
\end{figure*}

This is clearly a function of $(k_u, k_v)$, and nothing else. The multiplier $x_u$ is determined entirely by $u$'s degree and the $y_e$ are properties of starter packs, common to every pair. So, whatever made $u$ and $v$ co-members---that they are both linguists, both in London---is erased by the rewiring, and enters $\lambda_{uv}$ nowhere.
As a result, even if we sample the exact same pairs as we do to calculate the observed statistic, the practical effect is that rewiring re-sorts pairs into bins by degree. A low-degree pair thus has $\lambda_{uv} \approx 0$ and lands in the null's zero bin almost surely.
This means that only high-degree pairs have enough memberships to collide by chance in the reshuffling, so only they can populate the null's $s>0$ bins.
Furthermore, because our observable is monotonically increasing with degree, each pair then carries its own high similarity into whichever bin its degree assigns it, artificially inflating the similarities in low-$s$ bins.
The effect of this can be seen in Fig.~\ref{fig:bad_null}, in which we compare the observed curve to a configuration null generated using the \texttt{bicm} package~\citep{saracco_entropybased_2025,saracco_inferring_2017, saracco_randomizing_2015, vallarano_fast_2021}.
This choice of null causes an apparent threshold to appear, leading to a conclusion that sharing fewer than four SPs leads to users sharing \textit{less} common ground than would be expected by chance. This is, of course, not the case; it is an artifact of the chosen null model. While we do not perform an analysis on microcanonical configuration models, it is almost certain that this confound still holds in those cases as well.

To correct this, we simply skip the reshuffling and sample baseline pairs from the $s=0$ bin based on their joint degree. So, for each observed pair $(u,v)$ in bin $s = s_{uv}$, we record the joint degree profile $(k_u, k_v)$, coarsen degrees into deciles, and sample zero-overlap pairs from the same $(\mathrm{decile}_u, \mathrm{decile}_v)$ cell. This baseline then holds the expected joint degree in each bin fixed, and differs only in whether the users share any starter packs. We thus define the observed difference from the baseline as
\begin{equation}
    \begin{split}
        \Delta(s) \;=\; \underbrace{\mathbb{E}\big[\mathrm{sim} \mid k_u, k_v,\, s_{uv} = s\big]}_{\text{share } s \text{ SPs}} \\
        \;-\; \underbrace{\mathbb{E}\big[\mathrm{sim} \mid k_u, k_v,\, s_{uv} = 0\big]}_{\text{share none}}.
    \end{split}
\label{eq:delta-supp}
\end{equation}


\section{Appendix B: Sensitivity to the Vocabulary Cut}
\label{app:maxdf}

Our measure removes any term appearing in more than a fraction, $\mathrm{max\_df} = 0.4$, of users. This parameter, in part, licenses our description of the measure as capturing
\emph{marked} vocabulary rather than lexical overlap in general, as it allows us to capture the residual similarity between arbitrary users after the platform-wide register has hypothetically been excised. In this section, we analyze the sensitivity of our results to this free parameter by recomputing the full
pipeline at $\mathrm{max\_df} \in \{0.2, 0.3, 0.4, 0.5, 0.6\}$, holding all else fixed. Results are shown in Table~\ref{tab:maxdf}. 

\begin{table*}[t]
\centering
\small
\begin{tabular}{lrrrrrrrrrrrr}
\toprule
& & & \multicolumn{4}{c}{observed $\langle\mathrm{sim}_s\rangle$} & \multicolumn{2}{c}{matched $\langle\widetilde{\mathrm{sim}}_s\rangle$} & \multicolumn{2}{c}{$\Delta(s)$} & \multicolumn{2}{c}{ratio} \\
\cmidrule(lr){4-7} \cmidrule(lr){8-9} \cmidrule(lr){10-11} \cmidrule(lr){12-13}
max\_df & df ceiling &$|\mathcal{T}|$ & $s{=}0$ & $s{=}1$ & $s{=}4$ & $s{\ge}8$ & $s{=}1$ & $s{\ge}8$ & $s{=}1$ & $s{\ge}8$ & $s{=}1$ & $s{\ge}8$ \\
\midrule
$0.2$ & $44{,}159$ & $367{,}290$ & $0.015$ & $0.073$ & $0.103$ & $0.113$ & $0.031$ & $0.044$ & $0.042$ & $0.069$ & $2.38$ & $2.57$ \\
$0.3$ & $66{,}239$ & $367{,}854$ & $0.021$ & $0.090$ & $0.123$ & $0.135$ & $0.043$ & $0.060$ & $0.048$ & $0.075$ & $2.12$ & $2.26$ \\
$\mathbf{0.4}$ & $\mathbf{88{,}318}$ & $\mathbf{368{,}113}$ & $\mathbf{0.028}$ & $\mathbf{0.104}$ & $\mathbf{0.138}$ & $\mathbf{0.151}$ & $\mathbf{0.053}$ & $\mathbf{0.072}$ & $\mathbf{0.051}$ & $\mathbf{0.079}$ & $\mathbf{1.97}$ & $\mathbf{2.09}$ \\
$0.5$ & $110{,}398$ & $368{,}239$ & $0.033$ & $0.114$ & $0.149$ & $0.163$ & $0.061$ & $0.082$ & $0.053$ & $0.081$ & $1.87$ & $1.98$ \\
$0.6$ & $132{,}478$ & $368{,}299$ & $0.038$ & $0.122$ & $0.157$ & $0.171$ & $0.067$ & $0.089$ & $0.054$ & $0.081$ & $1.80$ & $1.91$ \\
\bottomrule
\end{tabular}
\caption{\textbf{Sensitivity to the vocabulary cut.} Mean observed similarity, degree-matched
baseline, degree-controlled effect $\Delta(s)$, and observed-to-matched ratio at
$\mathrm{max\_df} \in \{0.2, \ldots, 0.6\}$. The vectorizer is re-fitted at each setting over all
$220{,}796$ users with usable text; because \texttt{scikit-learn} interprets a fractional
$\mathrm{max\_df}$ as a proportion of the fitted corpus, we also report the corresponding
absolute document-frequency ceiling. The bolded row is the operating point used throughout the
main text.  Bin sizes are near-balanced from
$s{=}1$ to $s{=}7$ ($22{,}791$ to $24{,}221$ pairs); the pooled $s{\ge}8$ bin contains $3{,}667$.
Absolute levels rise by roughly a factor of two across the range, while $\Delta(s)$ remains
positive and monotone in $s$ and the ratio remains bounded well away from unity at every
setting.}
\label{tab:maxdf}
\end{table*}

As expected, this sensitivity analysis reveals that lowering $\mathrm{max\_df}$ prunes more vocabulary and leaves vectors composed of rarer terms, so cosine
similarities fall across the board. Similarly, raising it retains more platform-general vocabulary, which is
shared by construction, so they rise. Mean observed similarity at $s=1$ ranges from $0.073$ at
$\mathrm{max\_df}=0.2$ to $0.122$ at $\mathrm{max\_df}=0.6$, and the degree-matched baseline at the same
level from $0.031$ to $0.067$. Nevertheless, the findings of this study are not dependent on absolute, but rather relative magnitude. We find that these remain broadly intact. 

As shown in Table~\ref{tab:maxdf}, we find that across the range of $\mathrm{max\_df}$, both $\langle\mathrm{sim}_s\rangle$ and $\Delta(s)$ increase in $s$ without exception, preserving our finding of monotonic growth. Moreover, the ratio decreases with $\mathrm{max\_df}$, but remains roughly a factor of $2$ throughout---even reaching $2.57$ at $s \ge 8$ for low values of $\mathrm{max\_df}$---which suggests that our finding that SP co-membership contributes meaningfully to common ground remains intact. 

The marginal slope over $s \in [1,8]$ behaves likewise, varying between
$5.6 \times 10^{-3}$ and $6.95 \times 10^{-3}$ per SP across the range while remaining positive
and of the same order throughout.

\section{Appendix C: TF-IDF vs. Embeddings}\label{supp:not_embeddings}
A natural alternative would represent each user with a pretrained sentence embedding and take
cosine similarity in that space. We choose not to do this primarily because dense encoders are
trained to be invariant to precisely the property we wish to measure.

A sentence encoder's objective is to place paraphrases close together. This is what renders such models useful for retrieval, and it is exactly what we
must avoid. While indeed, some part of Clark's notion of common ground refers to the semantic content of two interlocutors' utterances, the dimension that this study is concerned with dives even deeper to assess the lexical resources they deploy to encode that information. Clark's canonical example is that while two people that are both from California might find themselves discussing matters related to the state, it is only if they are both natives of California and natives of San Francisco that they would be able to talk specifically about Chrissy Field~\citep{clark_context_2009}. An encoder which maps all conversations to within the same neighborhood is likely to miss this precise linguistic signal which our study relies on. Said otherwise, any particular lexical item is not a noisy encoding of some underlying concept which we would rather recover---rather, that particular lexical item \emph{is} the signal, and an encoder trained to discard it would
systematically erase that signal. Conversely, the failure would run in both
directions. Two users writing about unrelated topics in a similar register---both hedged,
both discursive, both academic in tone---will embed closer than two users writing about the same
niche subject in different registers, which cuts against a design targeting content rather than
style.

We do, however, employ pre-trained embeddings for starter pack names and descriptions (see Methods: Semantic Renormalization of Shared Starter Packs), where the task genuinely is precisely the inverse. In this case, it is the semantic similarity between starter packs titled, for example, ``New York City Cyclists'' and ``Peloton of NYC'' that we are trying to capture so that we can reduce redundancy in our signal of community co-membership.

\section{Appendix D: The Distinctness Gradient Is Not an Artifact of Person-Centric Starter Packs}
\label{app:personpacks}
 
A large class of starter packs is person-centric (e.g., ``Priya's Starter Pack''), and
these pose a specific hazard for the cross-cutting analysis. Because the naming
template dominates their embeddings, such SPs embed closer together (mean pairwise
cosine distance $\approx 0.48$) compared to topical SPs (mean pairwise cosine distance $\approx 0.79$),
so the renormalization collapses co-occurrence in several individuals' personal SPs
toward $s^{R} \approx 1$ regardless of the communities those SPs gather. Person-centric
SPs consequently concentrate at low $s^{R}$. A conservative pattern match on possessive SP names flags $31.8\%$ as person-centric, with these being more prevalent in lower $s^{R}$ bins. SP type and
semantic redundancy are thus entangled exactly where the profiles of
Fig.~\ref{fig:crosscutting} begin, and the rise of similarity with $s^{R}$ could, in
principle, reflect changing composition rather than distinctness.
 
Two observations rule this out. First, SP type is a small and approximately uniform
level offset. This is tested by classifying pairs by the composition of their shared packs, so that pairs are labeled as \emph{person-dominated} if at least half are possessive-named, \emph{topical} if none
are, and excluded if they are mixed. Comparing the two retained groups at fixed $(s, s^{R}\text{-bin})$ with
paired user-clustered bootstrap resamples, topical pairs are slightly more similar,
by roughly $+0.005\ [+0.0024, +0.0083]$ on average. In
the few strata where the difference resolves, this reaches $\approx +0.012$, but remains indistinguishable from zero in
most, exhibiting no concentration at low $s^{R}$. This uniform offset cannot generate the rise of similarity with $s^R$ seen in Fig.~\ref{fig:crosscutting}. Nevertheless, it exhibits some contribution to the effect, which we approximate to be $\approx 0.002$ of the observed
$\approx 0.05$ rise based on the $0.3$--$0.4$ decline in possessive share across
a profile, times a $\approx 0.005$ offset.
 
Second, the gradient we observe survives with these flagged packs removed entirely.
Table~\ref{tab:topical_spans} reports the results of re-running the same analysis on only the $51{,}126$ topical pairs as the mean across all $s$ bins.  
As observed, all differences between subsequent $s^R$ strata are positive, conserving the monotonic growth we witness in Fig.~\ref{fig:crosscutting},
with spans of $+0.019$ to $+0.055$ across $s \in [2,7]$ and intervals excluding zero in
all but $s = 7$, whose interval touches it. The pooled within-stratum span is $+0.034$
$[+0.023, +0.043]$, effectively the magnitude of the full analysis. 
 
\begin{table}[t]
\centering
\small
\begin{tabular}{llrr}
\toprule
$s$ & span (first $\to$ last bin) & estimate & $95\%$ CI \\
\midrule
$2$ & $\le 1.5 \to (1.5, 2.5]$ & $+0.022$ & $[+0.017, +0.026]$ \\
$3$ & $\le 1.5 \to (2.5, 3.5]$ & $+0.019$ & $[+0.008, +0.030]$ \\
$4$ & $\le 1.5 \to (3.5, 4.5]$ & $+0.055$ & $[+0.034, +0.075]$ \\
$5$ & $\le 1.5 \to (3.5, 4.5]$ & $+0.052$ & $[+0.023, +0.080]$ \\
$6$ & $(1.5, 2.5] \to (4.5, 5.5]$ & $+0.046$ & $[+0.024, +0.066]$ \\
$7$ & $(1.5, 2.5] \to {>}5.5$ & $+0.041$ & $[-0.008, +0.076]$ \\
\midrule
\multicolumn{2}{l}{pooled within-stratum span} & $+0.034$ & $[+0.023, +0.043]$ \\
\bottomrule
\end{tabular}
\caption{\textbf{The gradient on topical SPs alone.} For pairs whose shared SPs
contain no possessive-named SPs: the difference in mean cosine similarity between the
last and first readable $s^{R}$ bin of each raw-count stratum, with $95\%$ intervals
from paired user-clustered bootstrap resamples ($500$ resamples). There is insufficient data for the $s{=}8$ stratum
rests, so it is excluded.}
\label{tab:topical_spans}
\end{table}
 
\section{Appendix E: Renormalization Statistics}

\begin{figure}
    \centering
    \includegraphics[width=\linewidth]{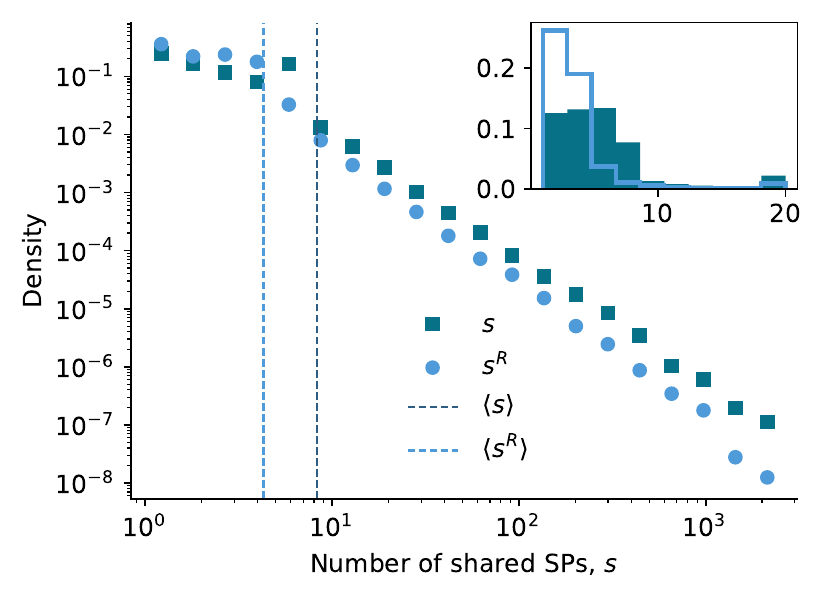}
    \caption{\textbf{Renormalizing the number of shared communities halves co-membership while preserving its heavy tail.} Here, we plot the distribution of raw ($s_{uv}$, square markers) and renormalized, or \textit{effective}, ($s^{R}_{uv}$, circle markers) numbers of shared starter packs. Dashed lines denote the mean values of the two distributions ($8.26$ and $4.29$, respectively). Inset: The two distributions for $s \in [0, 20]$, where the bulk of the differences occur. All values corresponding to $s > 20$ are pooled into the last bin.}
    \label{fig:distribution}
\end{figure}

Renormalization removes a great deal of nominal overlap, especially in the $s \in [0, 10]$ range. As a result, mean co-membership falls from $8.26$ SPs
to $4.29$ SPs, and the median from $5$ to $2.91$ (Fig.~\ref{fig:distribution}). In total, $87.7\%$ of co-member pairs have some redundancy
removed, and the median pair retains only $67\%$ of its nominal count. For $13.2\%$ of pairs
the effective count collapses to exactly $1$. The heavy tail nonetheless survives, remaining approximately power-law over
three orders of magnitude.

\section{Appendix F: Cosine Similarity at Higher Values of $s$}
\begin{figure}
    \centering
    \includegraphics[width=\linewidth]{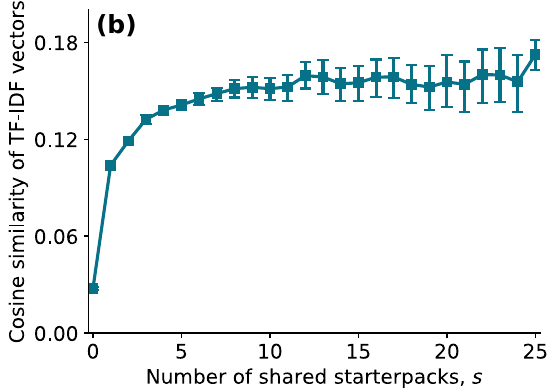}
    \caption{\textbf{Common ground remains stable into high $s$.} Observed cosine similarities extended to $s = 24$ with pooling at $s \geq 25$ suggest that the plateau in cosine similarity continues into the long tail of shared SPs.}
    \label{fig:extended}
\end{figure}

The long tail of $s$ witnessed in Fig.~\ref{fig:distribution} raises questions regarding the behavior of the cosine similarity measured as the number of shared SPs grows across orders of magnitude. Fig.~\ref{fig:extended} shows this up to $s=24$, with pooling at $s \geq 25$. From this, it is evident that changes in cosine similarity remain positive but marginal, and confidence intervals get increasingly large. Because of the lack of statistical power and relative monotonicity, we focus our results on the region $0 \leq s \leq 8$. Beyond $s = 24$, confidence intervals become too large to interpret. 
\end{document}